\documentclass[conference]{IEEEtran}
\usepackage{amsmath}
\usepackage{graphicx}
\usepackage{caption}
\usepackage{amsfonts}
\usepackage{color}
\usepackage{subcaption}
\usepackage{enumerate} 
\usepackage{dsfont}
\usepackage{amssymb}

\usepackage{comment}

\IEEEoverridecommandlockouts
\begin{document}
\title{MIMO-Zak-OTFS with Superimposed Spread Pilots}
\author{Abhishek Bairwa and Ananthanarayanan Chockalingam \\
Department of ECE, Indian Institute of Science, Bangalore}
\maketitle

\begin{abstract}  
In this paper, we consider the problem of spread pilot design and effective channel estimation in multiple-input multiple-output Zak-OTFS (MIMO-Zak-OTFS) with superimposed spread pilots, where data and spread pilot signals are superimposed in the same frame. To achieve good estimation performance in a MIMO setting, the spread pilots at different transmit antennas need to be effectively separated at the receiver. Towards this, we propose a spread pilot design that separates the pilot sequences in the cross-ambiguity domain and enables the estimation of the effective channel taps by a simple read-off operation. To further alleviate the effect of pilot-data interference on performance, we carry out turbo iterations between channel estimation and detection. Simulation results for $2\times 2$ and  $3\times 3$ MIMO-Zak-OTFS with Gaussian-sinc pulse shaping filter for vehicular-A channel model show that the proposed pilot design and estimation scheme with three turbo iterations can achieve very good estimation/detection performance. 
\end{abstract}
\vspace{1mm}

\section{Introduction}
\label{sec:Sec1}
Orthogonal time frequency space (OTFS) modulation, a delay-Doppler (DD) domain modulation, is effective in doubly-selective channels that are expected to be common in future wireless communication systems \cite{otfs_intro1},\cite{mc_otfs2},\cite{best_reads}. Recently, Zak transform-based OTFS (Zak-OTFS) modulation has been shown to be more robust to larger delay and Doppler spreads compared to multicarrier-based OTFS (MC-OTFS) \cite{otfs_book}-\cite{swaroop1}, In Zak-OTFS, information symbols are multiplexed in the DD domain and converted directly to a time domain (TD) signal for transmission using inverse Zak transform. A DD domain pulse shaping filter is used to limit the bandwidth and time duration of the transmitted signal \cite{embedded_zak},\cite{gauss_sinc}. At the receiver, channel estimation and data detection tasks are performed after time-to-DD domain conversion. Different types of pilot frames can be considered for channel estimation. They include: $(i)$ exclusive point pilot frame, where a frame is dedicated exclusively to a pilot symbol 
\cite{zak_otfs2}, $(ii)$ embedded point pilot frame, where a frame consists of pilot symbol(s) and data symbols with guard space in between to reduce pilot-data interference \cite{embedded_zak}, \cite{embedded_zak_2} and $(iii)$ superimposed spread pilot frame, where a spread pilot signal is superimposed on the data signal in the same frame \cite{superimposed_zak}. Exclusive and embedded point pilot frames incur throughput loss and have high peak-to-average power ratio (PAPR). In superimposed spread pilot frame, the spread pilot is obtained by passing a point pilot through a DD domain chirp filter, resulting in a uniform distribution of point pilot energy over the entire frame that reduces the PAPR \cite{superimposed_zak}. Also, since there is no guard space and the entire frame is used for data symbol multiplexing, there is no throughput loss. However, receiver processing with superimposed spread pilot gets more involved due to pilot-data interference. 

Past works on Zak-OTFS have considered the problem of channel estimation and detection in single-input single-output (SISO) settings. Multiple-input multiple-output Zak-OTFS (MIMO-Zak-OTFS) systems with multiple transmit (Tx) and receive (Rx) antennas are of interest to harness the rate and diversity benefits of MIMO. Recently, channel estimation/detection in MIMO-Zak-OTFS with exclusive and embedded point pilot frame has been considered in \cite{mimo_zak_exclusive} and \cite{mimo_zak_embedded}, respectively. In this paper, motivated by the throughput and PAPR advantages, we consider channel estimation/detection in Zak-OTFS with superimposed spread pilots in a MIMO setting, which has not been reported before.  

In MIMO-Zak-OTFS, the DD spread of the end-to-end effective channel between a pair of Tx and Rx antennas comprises of the DD spreads due to the Tx/Rx DD pulse shaping filters and the physical channel. The channel estimation problem here amounts to estimating the effective channel taps between all Tx/Rx antenna pairs. The effective channel taps between a pair of Tx/Rx antennas with superimposed spread pilot are estimated by computing cross-ambiguity of the received signal with the spread pilot signal. In a MIMO setting, different spread pilots are sent on different Tx antennas. In order to achieve good estimation in this scenario, the spread pilots from different Tx antennas need to be effectively separated at the receiver. Our first new contribution in this paper is that we propose a spread pilot design that separates the pilot sequences in the cross-ambiguity domain and enables the estimation of the effective channel taps by a simple read-off operation. Next, in order to further alleviate the effect of pilot-data interference on performance, we carry out turbo iterations between channel estimation and detection. Simulation results for $2\times 2$ and $3\times 3$ MIMO-Zak-OTFS with Gaussian-sinc pulse shaping filter and vehicular-A channel \cite{EVAITU} show that the proposed pilot design and estimation with three turbo iterations can achieve very good estimation/detection performance. 

\begin{figure*}[tp]
\centering
\vspace{0mm}
\includegraphics[width=17cm,height=9cm]{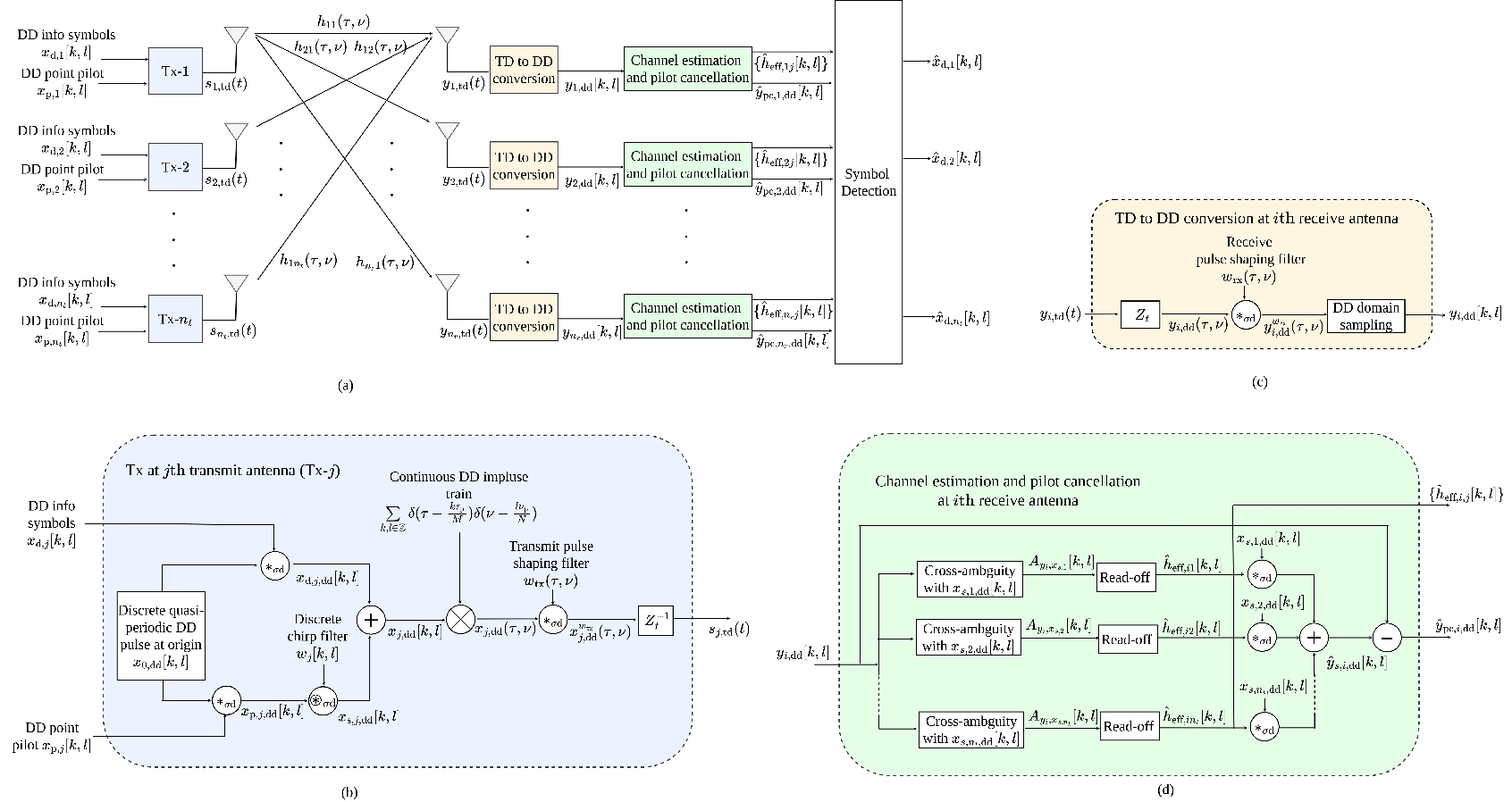}
\vspace{2mm}
\caption{(a) Block diagram of MIMO-Zak-OTFS transceiver with superimposed spread pilot. (b) Tx at $j\text{th}$ transmit antenna (Tx-$j$). (c) TD to DD conversion at $i\text{th}$ receive antenna. (d) Channel estimation and pilot cancellation at $i\text{th}$ receive antenna.} 
\label{fig:mimo_zak_otfs_block_diagram}
\vspace{-2mm}
\end{figure*}

\vspace{2mm}
\section{MIMO-Zak-OTFS system model with superimposed spread pilot}
\label{sec:Sec2}
In Zak-OTFS, the basic information carrier is a quasi-periodic DD domain pulse which is localized in the fundamental DD region $\mathcal{D}_{0}$ parameterized by delay period $\tau_{\text{p}}$ and Doppler period $\nu_{\text{p}}$, defined as $\mathcal{D}_0 = \{(\tau, \nu) \mid 0 \leq \tau < \tau_{\text{p}}, 0 \leq \nu < \nu_{\text{p}} \}$, where $\tau$ and $\nu$ are delay and Doppler variables, respectively. $\mathcal{D}_{0}$ is sliced into $M$ bins along the delay axis and $N$ bins along the Doppler axis to give the fundamental information grid ${\left\{\left(k\frac{\tau_{\text{p}}}{M},l\frac{\nu_{\text{p}}}{N}\right) \mid k=0, \cdots\hspace{-0.5mm}, M-1, l=0, \cdots\hspace{-0.5mm}, N-1 \right\}}$.

The block diagram of a MIMO-Zak-OTFS transceiver with superimposed spread pilot is shown in Fig. \ref{fig:mimo_zak_otfs_block_diagram}(a). There are $n_{t}$ transmitter antennas and $n_{r}$ receiver antennas. Figure \ref{fig:mimo_zak_otfs_block_diagram}(b) shows the transmit chain of the $j\text{th}$ transmit antenna. The discrete DD signal at the $j\text{th}$ transmitter is constructed using two parts: a data signal and a spread pilot signal. The discrete DD data signal is obtained by discrete twisted convolution\footnote{$a[k,l]*_{\sigma\text{d}}b[k,l] = \Sigma_{k',l'\in\mathbb{Z}}a[k-k',l-l']b[k',l']e^{j2\pi\frac{k'(l-l')}{MN}}.$} operation between the discrete DD function consisting of impulses at fundamental grid points mounted by $MN$ data symbols (i.e., $\Sigma_{k'= 0}^{M-1}\Sigma_{l'=0}^{N-1}x_{\text{d,j}}[k',l']\delta[k-k']\delta[l-l']$, where $\delta[.]$ is discrete Dirac-delta function) and the discrete quasi-periodic pulse at origin $x_{0,\text{dd}}[k,l] = \Sigma_{n,m \in \mathbb{Z}}\delta[k-nM]\delta[l-mN]$, given by

\vspace{-3mm}
{\small
\begin{eqnarray}
x_{\text{d,j,dd}}[k,l] & \hspace{-2mm} = & \hspace{-2mm} \frac{1} {\sqrt{MN}}\left(\Sigma_{k'=0}^{M-1}\Sigma_{l'=0}^{N-1}x_{\text{d,j}}[k',l']\delta{[k-k']}\delta{[l-l']}\right) \nonumber \\ 
& \hspace{-2mm} & \hspace{-2mm}  *_{\sigma\text{d}}\,x_{\text{0,dd}}[k,l] \nonumber \\ 
& \hspace{-2mm} = & \hspace{-2mm} \frac{1}{\sqrt{MN}}\Sigma_{k'=0}^{M-1}\Sigma_{l'=0}^{N-1}x_{\text{d,j}}[k',l']\Sigma_{n,m\in\mathbb{Z}}\delta[k-k'-nM] \nonumber \\
& \hspace{-2mm} & \hspace{-2mm} e^{j2\pi\frac{nl'}{N}}\delta[l-l'-mN],
\end{eqnarray}}

\hspace{-4.5mm}
where the factor $1/\sqrt{MN}$ is to ensure unit average energy for the discrete DD data signal. The spread pilot signal at the $j\text{th}$ transmitter is obtained as follows. First, a unit energy point pilot DD signal is obtained by discrete twisted convolution between the unit energy DD impulse function at $(k_{\text{p},j},l_{\text{p},j})$ and the discrete quasi-periodic pulse at origin $x_{\text{0,dd}}[k,l]$, given by $x_{\text{p},j,\text{dd}} = \delta[k-k_{\text{p},j}]\delta[l-l_{\text{p},j}] *_{\sigma\text{d}}x_{\text{0,dd}}[k,l]= \Sigma_{n,m\in\mathbb{Z}}e^{j2\pi\frac{nl_{\text{p},j}}{N}}\delta[k-k_{\text{p},j}-nM]\delta[l-l_{\text{p},j}-mN]$. An $MN$-periodic discrete chirp filter (i.e., $w_{j}[k,l] = \frac{1}{MN}e^{j2\pi\frac{q_{j}(k^2+l^2)}{MN}}$ where $q_{j}$ denotes the slope parameter) is then applied to $x_{\text{p},j,\text{dd}}[k,l]$. The action of the $MN$-periodic filter on DD point pilot signal is given by $MN$-periodic discrete twisted convolution\footnote{$a[k,l]\circledast_{\sigma\text{d}}b[k,l] = \Sigma_{k'=0}^{MN-1}\Sigma_{l'=0}^{MN-1}[k',l']b[k-k',l-l']e^{j2\pi\frac{l'(k-k')}{MN}}$.} (denoted by $\circledast_{\sigma\text{d}}$), and the resulting spread pilot signal is 
\begin{eqnarray}
\label{eq:spreadpilot}
x_{\text{s},j,\text{dd}}[k,l] & \hspace{-2mm} = & \hspace{-2mm} w_{j}[k,l]\circledast_{\sigma\text{d}}x_{\text{p},j,\text{dd}}[k,l] \nonumber \\
& \hspace{-2mm} = & \hspace{-2mm} \Sigma_{n=0}^{N-1}\Sigma_{m=0}^{M-1}w_{j}[k-k_{\text{p},j}-nM, l -l_{\text{p},j}-mN] \nonumber \\ 
& \hspace{-2mm} & \hspace{-2mm} e^{j2\pi\frac{nl_{\text{p},j}}{N}}e^{j2\pi\frac{(l-l_{\text{p},j}-mN)(k_{\text{p},j}+nM)}{MN}}.
\end{eqnarray}
The discrete superimposed DD signal at the $j\text{th}$ transmitter is obtained as $x_{j,\text{dd}}[k,l] = \sqrt{\frac{E_{\text{d}}}{n_{t}}} x_{\text{d},j,\text{dd}}[k,l]+ \sqrt{\frac{E_{\text{p}}}{n_{t}}}x_{\text{s},j,\text{dd}}[k,l]$, where $E_{\text{d}}$ and $E_{\text{p}}$ are the total average transmit data and pilot energies. The $x_{j,\text{dd}}[k,l]$s are supported on the information lattice $\Lambda_{\text{dd}}={\left\{\left(k\frac{\tau_{\text{p}}}{M},l\frac{\nu_{\text{p}}}{N}\right) \mid k,l\in\mathbb{Z} \right\}}$. The continuous DD superimposed signal is given by $x_{j,\text{dd}}(\tau,\nu) = \Sigma_{k,l\in\mathbb{Z}}x_{j,\text{dd}}[k,l]\delta(\tau-\frac{k\tau_{\text{p}}}{M})\delta(\nu-\frac{l\nu_{\text{p}}}{N})$, where $\delta(.)$ is continuous Dirac-delta function. A transmit DD domain pulse shaping filter $w_{\text{tx}}(\tau,\nu)$ is applied to $x_{j,\text{dd}}(\tau,\nu)$ to limit the transmission bandwidth and time duration\footnote{Commonly used DD domain pulse shaping filters are sinc, Gaussian, root raised cosine, and Gaussian-sinc filters. We consider Gaussian-sinc filter \cite{gauss_sinc}, given by $w_{\text{tx}}(\tau,\nu) = \Omega_{\tau}\Omega_{\nu}\sqrt{BT}\text{sinc}(B\tau)\text{sinc}(T\nu)e^{-\alpha_{\tau}B^{2}\tau^2}e^{-\alpha_{\nu}T^{2}\nu^2}$. To ensure no bandwidth/time expansion in Gaussian-sinc filter, $\alpha_{\tau} = \alpha_{\nu} = 0.044, \Omega_{\tau} = \Omega_{\nu}= 1.0278$ is used.}. The pulse shaped continuous DD domain signal is given by twisted convolution of $w_{\text{tx}}(\tau,\nu)$ with $x_{j,\text{dd}}(\tau,\nu)$ i.e., $x_{j,\text{dd}}^{w_{\text{tx}}}(\tau,\nu) = w_{\text{tx}}(\tau,\nu)*_{\sigma}x_{j,\text{dd}}(\tau,\nu)$, where $*_{\sigma}$ denotes the twisted convolution\footnote{$a(\tau,\nu)*_{\sigma}b(\tau,\nu) = \int\int a(\tau',\nu')b(\tau-\tau',\nu-\nu')e^{j2\pi\nu'(\tau-\tau')}d\tau'd\nu'.$}. The transmitted signal from $j\text{th}$ transmitter is the time domain (TD) realization of $x_{j,\text{dd}}^{w_{\text{tx}}}(\tau,\nu)$ given by $s_{j,\text{td}}(t) = Z_{t}^{-1}(x_{j,\text{dd}}^{w_{\text{tx}}}(\tau,\nu))$, where $Z_{t}^{-1}$ denotes the inverse Zak transform\footnote{$Z_{t}^{-1}(a(\tau,\nu)) =\sqrt{\tau_{\text{p}}}\int_{0}^{\nu_{\text{p}}}a(t,\nu)d\nu.$}.

The transmit TD signals $s_{j,\text{td}}(t)$, $j=1,\cdots,n_t$ pass through a doubly-selective channel. The physical DD channel between the $j\text{th}$ Tx antenna and $i\text{th}$ Rx antenna is given by $h_{ij}(\tau,\nu) = \Sigma_{p = 1}^{P}h_{ijp}\delta(\tau-\tau_{ijp})\delta(\nu-\nu_{ijp})$, where $P$ is the number of paths, and $h_{ijp}$, $\tau_{ijp}$, and $\nu_{ijp}$ are the fade coefficient, delay shift, and Doppler shift associated with $p\text{th}$ path, respectively. The received TD signal at $i\text{th}$ Rx antenna is given by $y_{i,\text{td}}(t) = r_{i,\text{td}}(t) + n_{i}(t) =  \Sigma_{j=1}^{n_{t}} \int\int h_{ij}(\tau,\nu) s_{j,\text{td}}(t - \tau)e^{j2\pi\nu(t-\tau)}d\tau d\nu + n_{i}(t),$ where $n_{i}(t)$ is AWGN with variance $N_{0}$. The average noise energy at the $i{\text{th}}$ receiver is $MNN_{0}$, so the data SNR $\rho_{\text{d}} = E_{\text{d}}/MNN_{0}$ and the pilot SNR $\rho_{\text{p}} = E_{\text{p}}/MNN_{0}$. The pilot-to-data ratio (PDR) is defined as the ratio of pilot SNR to data SNR, i.e., PDR = $\frac{\rho_{\text{p}}}{\rho_{\text{d}}} = \frac{E_{\text{p}}}{E_{\text{d}}}$. 

As shown in Fig.\ref{fig:mimo_zak_otfs_block_diagram}(c), the TD signal at the $i\text{th}$ Tx antenna is converted to DD domain via Zak transform\footnote{$Z_{t}(a(t)) = \sqrt{\tau_{\text{p}}}\Sigma_{k\in\mathbb{Z}}a(\tau+k\tau_{\text{p}})e^{-j2\pi\nu k\tau_{\text{}p}}$.}, i.e., $y_{i,\text{dd}}(\tau,\nu) = Z_{t}(y_{i,\text{td}}(t)) = r_{i,\text{dd}}(\tau,\nu) + n_{i,\text{dd}}(\tau,\nu)$, where $r_{i,\text{dd}}(\tau,\nu) = \Sigma_{j=1}^{n_{t}}h_{ij}(\tau,\nu)*_{\sigma}w_{\text{tx}}(\tau,\nu)*_{\sigma}x_{j,\text{dd}}(\tau,\nu)$ is the Zak transform of $r_{i,\text{td}}(t)$ and $n_{i,\text{dd}}(\tau,\nu)$ is the Zak transform of $n_{i}(t)$. The receive DD filter $w_{\text{rx}}(\tau,\nu)$ matched to the Tx DD filter acts on $y_{i,\text{dd}}(\tau,\nu)$ through twisted convolution to give output $y_{i,\text{dd}}^{w_{\text{rx}}}(\tau,\nu) = w_{\text{rx}}(\tau,\nu)*_{\sigma}y_{i,\text{dd}}(\tau,\nu) = \Sigma_{j = 1}^{n_{t}}h_{\text{eff},ij}(\tau,\nu)*_{\sigma}x_{j,\text{dd}}(\tau,\nu) + n_{i,\text{dd}}^{w_{\text{rx}}}(\tau,\nu)$, where $h_{\text{eff},ij}(\tau,\nu) = w_{\text{rx}}(\tau,\nu)*_{\sigma}h_{ij}(\tau,\nu)*_{\sigma}w_{\text{tx}}(\tau,\nu)$ is the effective channel between $i\text{th}$ receiver and $j\text{th}$ transmitter, and $n_{i,\text{dd}}^{w_{\text{rx}}}(\tau,\nu) = w_{\text{rx}}(\tau,\nu)*_{\sigma}n_{i,\text{dd}}(\tau,\nu)$, is the filtered DD noise at the $i\text{th}$ receiver. The continuous DD signal $y_{i,\text{dd}}^{w_{\text{rx}}}(\tau,\nu)$ is sampled on the information lattice $\Lambda_{\text{dd}}$ which results in discrete DD signal $y_{i,\text{dd}}[k,l] = y_{i,\text{dd}}^{w_{\text{rx}}}(\tau = \frac{k\tau_{\text{p}}}{M},\nu = \frac{l\nu_{\text{p}}}{N}) = \Sigma_{j=1}^{n_{t}}h_{\text{eff},ij}[k,l]*_{\sigma\text{d}}x_{j,\text{dd}}[k,l] + n_{\text{i,dd}}[k,l]$, where $h_{\text{eff},ij}[k,l]*_{\sigma\text{d}}x_{j,\text{dd}}[k,l] = \Sigma_{k',l'\in\mathbb{Z}}h_{\text{eff},ij}[k-k',l-l']x_{j,\text{dd}}[k',l']e^{j2\pi\frac{k'(l-l')}{MN}}$, the filtered noise samples $n_{i,\text{dd}}[k,l] = n_{i,\text{dd}}^{w_{\text{rx}}}(\tau=\frac{k\tau_{\text{p}}}{M},\nu=\frac{l\nu_{\text{p}}}{N})$, and the effective filter channel taps $h_{\text{eff},ij}[k,l] = h_{\text{eff},ij}(\tau=\frac{k\tau_{\text{p}}}{M},\nu = \frac{l\nu_{\text{p}}}{N})$. As $y_{i,\text{dd}}[k,l]$ and $x_{j,\text{dd}}[k,l]$ are quasi-periodic, it is sufficient to consider the samples in the fundamental grid. 

We can write $y_{i,\text{dd}}[k,l]$ and $x_{j,\text{dd}}[k,l]$ in vectorized form, resulting in a vectorized signal model 
$\mathbf{y}_{i} = \Sigma_{j=1}^{n_{t}}\mathbf{H}_{ij}\mathbf{x}_{j} + \mathbf{n}_{i}$, where $\mathbf{y}_{i},\mathbf{x}_{j},\mathbf{n}_{i}\in \mathbb{C}^{MN\times 1}$, such that their $(kN+l+1)\text{th}$ element is given by $(\mathbf{x}_{j})_{kN+l+1} = x_{j,\text{dd}}[k,l]$, $(\mathbf{y}_{i})_{kN+l+1} = y_{i,\text{dd}}[k,l]$, $(\mathbf{n}_{i})_{kN+l+1} = n_{i,\text{dd}}[k,l]$, and $\mathbf{H}_{ij}\in\mathbb{C}^{MN\times MN}$ such that $\mathbf{H}_{ij}[k'N+l'+1,kN+l+1] = \Sigma_{n,m \in \mathbb{Z}}h_{\text{eff},ij}[k'-k-nM, l'-l-mN]e^{j2\pi\frac{nl}{N}}e^{j2\pi\frac{(l'-l-mN)(k+nM)}{MN}}$, where $k',k = 0,1,\cdots ,M-1, l',l = 0 ,1, \cdots, N-1$. These $\mathbf{y}_{i}$ vectors for $i = 1,\cdots,n_{t}$ can be concatenated to obtain the vectorized MIMO input-output (I/O) relation as
\small
\begin{align} 
\underbrace{
\begin{bmatrix}
\mathbf{y} _{1}    \\
\mathbf{y} _{2} \\ \vdots \\  
\mathbf{y} _{n_{r}}
\end{bmatrix}
}_{{\bf y}_{\tiny \mbox{MIMO}}} & = 
\underbrace{
\begin{bmatrix}
\mathbf{H} _{11} & \mathbf{H} _{12} & \dots &  \mathbf{H} _{1n_{t}} \\
\mathbf{H} _{21} & \mathbf{H} _{22} & \dots &  \mathbf{H} _{2n_{t}} \\
\vdots & \vdots & \ddots &\vdots\\
\mathbf{H} _{n_{r}1} & \mathbf{H} _{n_{r}2} & \dots &  \mathbf{H} _{n_{r}n_{t}} \\
\end{bmatrix} 
}_{{\bf H}_{\tiny \mbox{eff,MIMO}}}
\underbrace{
\begin{bmatrix}
\mathbf{x} _{1}    \\
\mathbf{x} _{2} \\ \vdots \\  
\mathbf{x} _{n_{t}}
\end{bmatrix}
}_{{\bf x}_{\tiny \mbox{MIMO}}}
+
\underbrace{
\begin{bmatrix}
\mathbf{n} _{1}    \\
\mathbf{n} _{2} \\ \vdots \\  
\mathbf{n} _{n_{r}}
\end{bmatrix}
}_{{\bf n}_{\tiny \mbox{MIMO}}},
\end{align}
\normalsize
where $\mathbf{y}_{\tiny\mbox{MIMO}},\mathbf{n}_{\tiny\mbox{MIMO}}\in \mathbb{C}^{n_{r}MN\times 1}$, $\mathbf{x}_{\tiny\mbox{MIMO}} \in \mathbb{C}^{n_{t}MN\times 1}$, and $\mathbf{H}_{\tiny\mbox{eff,MIMO}}\in\mathbb{C}^{n_rMN \times n_{t}MN}$. The Rx antennas are assumed to be placed sufficiently apart so that the noise vectors are independent of each other. Thus, $\mathbf{C}_{\mathbf{n}_{\tiny\mbox{MIMO}}} = \mathbb{E}[\mathbf{n}_{\tiny\mbox{MIMO}}\mathbf{n}^{H}_{\tiny\mbox{MIMO}}] = \mathit{I}_{n_{r}}\otimes \mathbf{C_{n}}$ where $\mathbf{C}_{\mathbf{n}} = \mathbf{C}_{\mathbf{n}_{i}} = \mathbb{E}[\mathbf{n}_{i}\mathbf{n}^{H}_{i}]$, $i = 1,\cdots,n_{r}$. The $y_{i,\text{dd}}[k,l]$s are further processed to estimate the effective MIMO channel matrix $\mathbf{H}_{\tiny\mbox{eff,MIMO}}$ and detect the data symbols.

\section{Proposed spread pilot construction}
\label{sec:construction}
The channel estimation task is to estimate the elements of the $\mathbf{H}_{\tiny\mbox{eff,MIMO}}$ matrix. For good estimation performance, the spread pilots at different Tx antennas need to be separated at the receiver side. In this section, we propose a construction that separates the spread pilots in the cross-ambiguity domain. The cross-ambiguity\footnote{Cross-ambiguity function between two discrete DD domain signals $a[k,l]$ and $b[k,l]$ is given by $A_{a,b}[k,l] = \Sigma_{k'=0}^{M-1}\Sigma_{l'=0}^{N-1}a[k',l']b^{*}[k'-k,l'-l]e^{-j2\pi\frac{l(k'-k)}{MN}}.$} $A_{y_{i},x_{s,j}}[k,l]$, between the received discrete DD signal at $i\text{th}$ Rx antenna  $y_{i,\text{dd}}[k,l]$ and the spread pilot signal at $j\text{th}$ Tx antenna $x_{s,j,\text{dd}[k,l]}$ is given by 

\vspace{-2mm}
{\small
\begin{eqnarray}
\label{cross_ambiguity_eq}
\hspace{-7mm}
A_{y_{i},x_{s,j}}[k,l] & \hspace{-2mm} = & \hspace{-2mm} \sqrt{\frac{E_{\text{p}}}{n_{t}}}h_{\text{eff},ij}[k,l]*_{\sigma_{{\text{d}}}}A_{x_{s,j},x_{s,j}}[k,l] \nonumber \\
& \hspace{-25mm} & \hspace{-20mm} + \sum\limits_{\substack{v=1 \\ v \neq j}}^{n_t}\sqrt{\frac{E_{\text{p}}}{n_{t}}}h_{\text{eff},iv}[k,l]*_{\sigma_{\text{d}}}A_{x_{s,v},x_{s,j}}[k,l] \nonumber \\
& \hspace{-25mm} & \hspace{-20mm} + \sum\limits_{v=1}^{n_{t}}\sqrt{\frac{E_{d}}{n_{t}}}h_{\text{eff},iv}[k,l]*_{\sigma_{\text{d}}}A_{x_{d,v},x_{s,j}}[k,l] 
+ A_{n_{i},x_{s,j}}[k,l],
\end{eqnarray}}

\hspace{-4.5mm}
where $A_{x_{s,j},x_{s,j}}$ is the self-ambiguity of the spread pilot at the $j\text{th}$ Tx, and $A_{x_{s,v},x_{s,j}}$, $A_{x_{d,v},x_{s,j}}$, and $A_{n_{i},x_{s,j}}$ are the cross-ambiguity between the spread pilot at the $j\text{th}$ Tx and $i)$ the spread pilot at the $v\text{th}$ Tx, $ii)$ data signal at the $v\text{th}$ Tx, and $iii)$ noise at the $i\text{th}$ Rx, respectively. For odd primes $M$,$N$ and $q_{j}$ being relatively prime to both $M$ and $N$, the self-ambiguity $A_{x_{s,j},x_{s,j}}$ is supported on a twisted lattice $\Lambda_{\text{q}_{j}}$ and expressed as $A_{x_{s,j},x_{s,j}}[k,l] = \Sigma_{k_{l},l_{l}\in\Lambda_{q_{j}}}e^{j\theta_{l}}\delta[k-k_{l}]\delta[l-l_{l}]$, where $\theta_{l}$ is the phase associated with the lattice point $(k_{l},l_{l})$ with $\theta_{l} = 0$ for $(k_{l},l_{l})=(0,0)$. The $(k_{l},l_{l})\in\Lambda_{q_{j}}$ satisfies 

\vspace{-1.5mm}
{\small
\begin{align}
(1-2q_{j}\theta_{j})
\begin{bmatrix}
\ k_{l}    \\
\ l_{l} \\  
\end{bmatrix}
=
\begin{bmatrix}
\ \theta_{j} & 1    \\
\ 1 & 2q_{j} \\  
\end{bmatrix}
\begin{bmatrix}
\ nM    \\
\ mN \\  
\end{bmatrix},
\end{align}}

\vspace{-1mm}
\hspace{-4.5mm}
where $\theta_{j} = [(2q_{j})^{-1 }-2q_{j}]_{MN}$, $(2q_{j})^{-1}$ is the unique inverse of $2q_{j}$ modulo $MN$ and $[.]_{MN}$ denotes modulo-$MN$ operation \cite{superimposed_zak}. Thus, the twisted lattice $\Lambda_{q_{j}}$ is obtained by a rotation of the period lattice $\Lambda_{\text{p}}$, i.e., ${\left\{\left(nM,mN\right)\mid n,m\in \mathbb{Z}\right\}}$.
Substituting the expression for $A_{x_{s,j},x_{s,j}}[k,l]$ in (\ref{cross_ambiguity_eq}), we obtain

\vspace{-4mm}
{\footnotesize
\begin{eqnarray}
\label{interference_cross_ambiguity_eq}
A_{y_{i},x_{s,j}}[k,l] & \hspace{-2mm = & \hspace{-2mm}  \underbrace{\sqrt{\frac{E_{\text{p}}}{n_{t}}}h_{\text{eff},ij}[k,l]}_{1.\, \text{Eff. chl. between $j\text{th}$ Tx antenna and $i\text{th}$ Rx antenna  }} \nonumber \\
&\hspace{-30mm} & \hspace{-25mm} + \underbrace{\sum\limits_{\substack{(k_{l},l_{l})\in \Lambda_{q}\\(k_{l},l_{l}) \neq (0,0)}}\hspace{-2.5mm}\sqrt{\frac{E_{\text{p}}}{n_{t}}}e^{j\theta_{l}}h_{\text{eff},ij}[k-k_{l},l-l_{l}]e^{j2\pi\frac{(l-l_{l})k_{l}}{MN}}}_{2.\,\text{DD aliasing due to spread pilot of }j\text{th Tx antenna}} \nonumber \\
& \hspace{-30mm} & \hspace{-25mm} + \underbrace{\sum\limits_{\substack{v=1 \\ v \neq j}}^{n_t}\sqrt{\frac{E_{\text{p}}}{n_{t}}}h_{\text{eff},iv}[k,l]*_{\sigma_{\text{d}}}A_{x_{s,v},x_{s,j}}[k,l]}_{3.\,\text{Interference due to spread pilots of other Tx antennas}} \nonumber \\
& \hspace{-30mm} & \hspace{-25mm} + \underbrace{\sum\limits_{v=1}^{n_{t}}\sqrt{\frac{E_{d}}{n_{t}}}h_{\text{eff},iv}[k,l]*_{\sigma_{\text{d}}}A_{x_{d,v},x_{s,j}}[k,l]}_{\text{4.\,Data interference from all Tx antennas}} 
+ \underbrace{A_{n_{i},x_{s,j}}[k,l]}_{5.\,\text{Receiver noise}}.
\end{eqnarray}
}

\normalsize
\hspace{-4.5mm}
In (\ref{interference_cross_ambiguity_eq}), $A_{y_{i},x_{s,j}}[k,l]$ has $5$ terms: the true effective channel  between $j\text{th}$ Tx and $i\text{th}$ Rx ($1\text{st}$ term), DD aliasing channel taps centered around self ambiguity lattice points other than origin ($2\text{nd}$ term), interference due to spread pilots at transmitters other than  $j\text{th}$ Tx ($3\text{rd}$ term), data interference from all transmitters ($4\text{th}$ term), and noise ($5\text{th}$ term).
\begin{figure}
\hspace{-8mm}
\includegraphics[width=9.5cm,height=7.5cm]{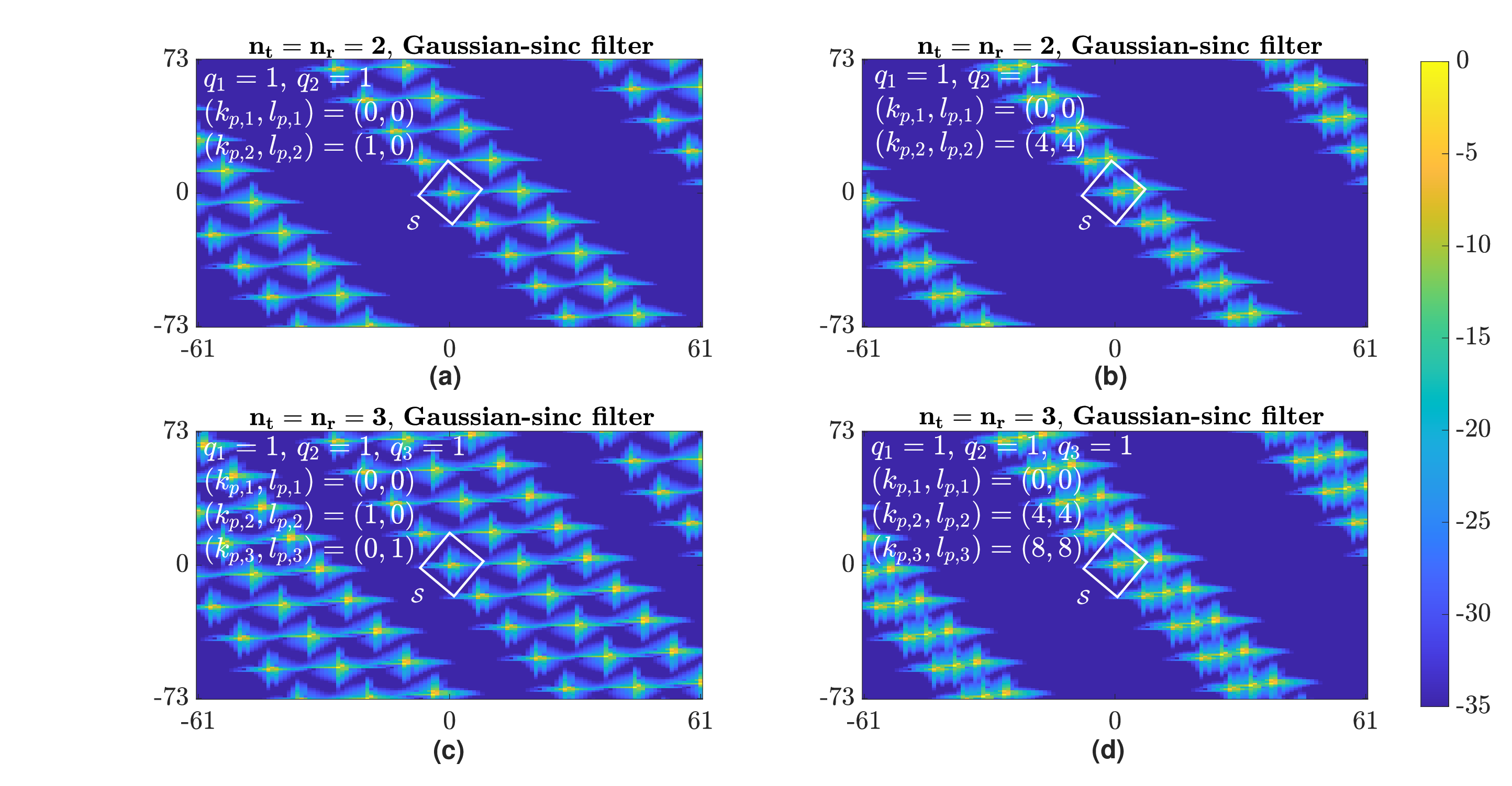}
\caption{Heatmaps of the cross-ambiguity
$A_{y_{1},x_{s,1}}[k,l]$ for $2\times2 $ and $3\times3 $ MIMO-Zak-OTFS with Gaussian-sinc filter.} 
\label{fig:cross_amb_plots}
\vspace{-10mm}
\end{figure}
Figure \ref{fig:cross_amb_plots} shows the heatmaps of the cross-ambiguity $A_{y_{1},x_{s,1}}[k,l]$ in the absence of data and noise (i.e., no $4\text{th}$ and $5\text{th}$ term in (\ref{interference_cross_ambiguity_eq})) for $2\times2$ and $3\times3$ MIMO systems. For a well-localized filter like Gaussian-sinc filter and carefully chosen slope parameter, the DD aliasing due to spread pilot at $1\text{st}$ Tx (2nd term in (\ref{interference_cross_ambiguity_eq})) can be less. To observe the interference due to spread pilots from other transmitters (3rd term in (\ref{interference_cross_ambiguity_eq})), two configurations are considered. In Fig. \ref{fig:cross_amb_plots}(a), we consider $2\times2$ MIMO with spread pilot at $1\text{st}$ Tx, i.e., $x_{s,1}[k,l]$, obtained by a chirp filter with slope parameter $q_{1} = 1$ to a point pilot located at $(k_{\text{p},1},l_{\text{p},1}) = (0,0)$. Similarly, $x_{s,2}[k,l]$ is obtained by passing a point pilot located at $(k_{\text{p},2},l_{\text{p},2}) = (1,0)$ through the chirp filter with same slope parameter, i.e., $q_{2} = q_{1} =1$. According to (\ref{interference_cross_ambiguity_eq}), $A_{y_{1},x_{s,1}}[k,l]$ consists of the following: $(i)$ true $h_{\text{eff},11}[k,l]$ taps centered around origin, $(ii)$ DD aliases of $h_{\text{eff},11}[k,l]$ taps centered around twisted lattice points of self-ambiguity function $A_{x_{s,1},x_{s,1}}[k,l]$ other than origin, and $(iii)$ $h_{\text{eff},12}[k,l]$ taps aliases centered around the points on which $A_{x_{s,2},x_{s,1}}[k,l]$ is supported. These points constitute a lattice which is twisted as well as shifted compared to the period lattice $\Lambda_{\text{p}}$. Due to the shift with respect to $\Lambda_{\text{p}}$, it is not centered at origin. This shift allows for estimation of $h_{\text{eff},11}[k,l]$ taps by a read-off operation inside the region $\mathcal{S}$ centered around origin. The interference due to $(iii)$ term inside $\mathcal{S}$ can be made less by increasing this shift. We observe that the shift depends upon the point pilot locations $(k_{\text{p},1},l_{\text{p},1})$, $(k_{\text{p},2},l_{\text{p},2})$ and the slope parameter $q_{1}$. In Fig. \ref{fig:cross_amb_plots}(b), we consider a different configuration with $(k_{\text{p},1},l_{\text{p},1}) = (0,0)$, $(k_{\text{p},2},l_{\text{p},2}) = (4,4)$ and $q_{1} = q_{2} = 1$. We observe that for this configuration, the interference due to $(iii)$ term inside $\mathcal{S}$ is significant and the read-off operation will result in a worse estimation. A similar illustration is shown for $3\times 3$ MIMO in Figs. \ref{fig:cross_amb_plots}(c) and \ref{fig:cross_amb_plots}(d). We formalize the observations above in the following theorem.

\textit{\bf Theorem 1.} 
Let $x_{s,j,\text{dd}}[k,l]$ and $x_{s,v,\text{dd}}[k,l]$ denote the DD spread pilot signals at $j\text{th}$ and $v\text{th}$ Tx antennas, respectively. Let $x_{s,j,\text{dd}}[k,l]$ be constructed by passing a point pilot at $(k_{\text{p},j},l_{\text{p,j}})$ through a chirp filter with slope parameter $q_{j}$ and let $x_{s,v,\text{dd}}[k,l]$ be constructed by applying a chirp filter with slope parameter $q_{v}$ to a point pilot at $(k_{\text{p},v},l_{\text{p},v})$. If $M,N$ are odd primes, $q_{j}\equiv q_{v}$ (mod $MN$)
and $q_{j}$ is relatively prime to both $M$ and $N$, then

\vspace{-4mm}
{\footnotesize
\begin{eqnarray}
\label{cross_ambiguity_spreadpilots}
|A_{x_{s,v},x_{s,j}}[k,l]| = 
\begin{cases}
1, & \hspace{0mm} 2q_{j}(k+k_{\text{p},j}-k_{\text{p},v})-l \equiv 0 \hspace{-2mm} \pmod{M}, \\
& \hspace{0mm}\theta_{j}l-2q_{j}(l_{\text{p},j}-l_{\text{p},v})-k\equiv0 \hspace{-2mm} \pmod{N},\\ 
0, & \hspace{0mm} \mbox{\small{otherwise,}} 
\end{cases}
\hspace{-3mm}
\end{eqnarray}}

\hspace{-4.5mm}
where $\theta_{j} \equiv (2q_{j})^{-1}-2q_{j} \pmod{MN}$, $(2q_{j})^{-1}$ is the unique inverse of $2q_{j}$ modulo $MN$. Also, $A_{x_{s,v},x_{s,j}}[k,l]$ is supported on a shifted lattice $\Lambda_{q_{j},\text{shift}}$ and the shifted lattice points $(k_{l_{\text{shift}}},l_{l_{\text{shift}}})$ are given by (the arithmetic is interpreted as modulo $MN$)

\vspace{-4mm}
{\small
\begin{eqnarray}
\hspace{-7mm}
(1-2q_{j}\theta_{j})
\begin{bmatrix}
\ k_{l_{\text{shift}}}    \\
\ l_{l_{\text{shift}}} \\  
\end{bmatrix}
& \hspace{-2mm} = & \hspace{-2mm}
\begin{bmatrix}
\ \theta_{j} & 1    \\
\ 1 & 2q_{j} \\  
\end{bmatrix}
\begin{bmatrix}
\ nM    \\
\ mN \\ 
\end{bmatrix}
\nonumber \\
& \hspace{-15mm} & \hspace{-10mm} + \begin{bmatrix}
\ 2q_{j}\theta_{j}(k_{\text{p,j}}-k_{\text{p},v})-2q_{j}(l_{\text{p},j}-l_{\text{p},v})    \\
\ -4q_{j}^{2}(l_{\text{p},j}-l_{\text{p},v})+2q_{j}(k_{\text{p},j}-k_{\text{p},v})\\ 
\end{bmatrix}.
\end{eqnarray}
} 

\begin{IEEEproof}
See Appendix. 
\end{IEEEproof} 

From the above theorem, it follows that, if $(k_{\text{p},j},l_{\text{p},j})\neq (k_{\text{p},v},l_{p,v})$, then the shifted lattice $\Lambda_{q_{j},\text{shift}}$ is  obtained by rotating and shifting the period lattice $\Lambda_{\text{p}}.$ This shifted lattice is not centered at origin, unlike the twisted lattice of the self-ambiguity function $A_{x_{s,j},x_{s,j}}[k,l]$. Therefore, the interference due to undesired channel taps centered on shifted lattice points of the cross-ambiguity function $A_{x_{s,v},x_{s,j}}[k,l]$ can be made less, allowing the desired effective channel taps centered at $(0,0)$ in the twisted lattice of $A_{x_{s,j},x_{s,j}}[k,l]$ to be read-off as the estimates. This can be done by choosing $(k_{\text{p},j},l_{\text{p},j}), (k_{\text{p},v},l_{p,v})$, $q_{j}$, $q_{v}$  appropriately. We have chosen appropriate parameters for $2\times2$ and $3\times3$ MIMO in Sec. \ref{sec:results}.  

\section{Channel estimation and signal detection}
\label{sec: chl. est.}
Using the pilot construction in Sec. \ref{sec:construction} that separates the spread pilots, we can obtain the estimate $\hat{h}_{\text{eff},ij}[k,l]$ taps by the following simple read-off\footnote{Not all taps of $\sqrt{(n_{t}/E_{\text{p}})}A_{y_{i},x_{s,j}}[k,l]$ are genuine taps of $h_{\text{eff},i,j}[k,l]$ due to randomness of noise and interference due to data. Only those taps are considered for which $\sqrt{(n_{t}/E_{\text{p}})}|A_{y_{i},x_{s,j}}[k,l]|$ exceeds 3 times the standard deviation of $(h_{\text{eff},i,j}[k,l] - \hat{h}_{\text{eff},i,j}[k,l])$. The variance of this quantity is $\frac{1}{MN}\left(\frac{1+\rho_{\text{d}}}{\rho_{\text{p}}.}\right)$} operation: 

\vspace{-2mm}
{\small
\begin{eqnarray}
\label{estimation_readoff}
\hat{h}_{\text{eff},ij}[k,l] = 
\begin{cases}
\sqrt{\frac{n_{t}}{E_{\text{p}}}}A_{y_{i},x_{s,j}}[k,l], & \hspace{4mm} (k,l)\in \mathcal{S}, \\
0, & \hspace{4mm} \mbox{\small{otherwise.}} \\
\end{cases}
\end{eqnarray}}

\vspace{-2mm}
\hspace{-4,5mm}
where $\mathcal{S}$ is the support region considered for estimation.
Using $\hat{h}_{\text{eff},ij}[k,l]$, $\mathbf{\hat{H}}_{ij}$ can be constructed. The $\mathbf{\hat{H}}_{ij}$s are concatenated to obtain $\mathbf{\hat{H}_{\tiny\mbox{eff,MIMO}}}$. The estimated $\hat{h}_{\text{eff},ij}[k,l]$ taps are used to reconstruct the received spread pilot signal at the $i{\text{th}}$ receiver i.e., $\hat{y}_{\text{s},i,\text{dd}}[k,l]=\Sigma_{j=1}^{n_{t}}\hat{h}_{\text{eff},i,j}[k,l]*_{\sigma\text{d}}\left(\sqrt{E_{\text{p}}/n_{t}}x_{\text{s},j,\text{dd}}[k,l]\right)$, and then subtract it from received signal at the $i\text{th}$ receiver to obtain
$\hat{y}_{\text{pc},i,\text{dd}}[k,l] = y_{i,\text{dd}}[k,l] - \hat{y}_{\text{s},i,\text{dd}}[k,l]$, as shown in Fig.\ref{fig:mimo_zak_otfs_block_diagram}(d). 
This pilot-cancelled signal $\hat{y}_{\text{pc},i,\text{dd}}[k,l]$ for $i=1,\cdots,n_{r}$ is used to detect the data symbols $x_{\text{d},j}[k,l]$ for $j = 1,\cdots,n_{t}$ using the vectorized I/O relation in Sec. \ref{sec:Sec2}. In this work, we use MMSE-LAS algorithm in \cite{mimo_book} for data detection, which is a iterative algorithm. Starting with the initial vector obtained using MMSE equalizer, the algorithm searches the local neighborhood and obtains a local minima as the solution. The performance can be further improved using turbo iterations between channel estimation and detection as follows. The $t\text{th}$ turbo iteration consists of following steps. \\
\vspace{1mm}
\textit{\textbf{Step 1:}}
The estimated channel taps and detected symbols in the $(t-1){\text{th}}$ iteration are used to reconstruct the received data signal at the $i\text{th}$ receiver as {\small  $\hat{y}_{\text{d},i,\text{dd}}^{(t)}[k,l] = \Sigma_{j=1}^{n_{t}}\hat{h}_{\text{eff},i,j}^{(t-1)}[k,l]*_{\sigma\text{d}}\left(\sqrt{E_{\text{d}}/n_{t}}\hat{x}_{\text{d},j,\text{dd}}^{(t-1)}[k,l]\right)$}, where $\hat{x}_{\text{d},j,\text{dd}}^{(t-1)}[k,l] = \Sigma_{k'=0}^{M-1}\Sigma_{l'=0}^{N-1}\hat{x}_{\text{d},j}^{(t-1)}[k',l']\delta[k-k']\delta[l-l']*_{\sigma\text{d}}x_{\text{0,dd}}[k,l]$. The data-cancelled received signal at the $i\text{th}$ receiver is obtained as {\small $\hat{y}_{\text{dc},i,\text{dd}}^{(t)}[k,l] = y_{i,\text{dd}}[k,l] - \hat{y}_{\text{d},i,\text{dd}}^{(t)}[k,l]$.} \\ 
\vspace{1mm}
\textit{\textbf{Step 2:}}
The channel taps in the $t{\text{th}}$ iteration is given by {\small $\hat{h}_{\text{eff},i,j}^{(t)}[k,l] = A_{\hat{y}_{\text{dc},i}^{(t)},x_{s,j}}[k,l]/\sqrt{E_{\text{p}}/n_{t}}$} for {\small $(k,l)\in\mathcal{S}$}, where {\small $A_{\hat{y}_{\text{dc},i}^{(t)},x_{s,j}}$} 
is the cross-ambiguity of {\small $\hat{y}_{\text{dc},i,\text{dd}}^{(t)}[k,l]$} with {\small $x_{\text{s},j,\text{dd}}[k,l]$.} \\
\vspace{1mm}
\textit{\textbf{Step 3:}}
The received spread pilot signal is reconstructed using the channel estimates as {\small $\hat{y}_{\text{s},i,\text{dd}}^{(t)}[k,l] =  \Sigma_{j=1}^{n_{t}}\hat{h}_{\text{eff},i,j}^{(t)}[k,l]*_{\sigma\text{d}}\left(\sqrt{E_{\text{p}}/n_{t}}\hat{x}_{\text{s},j,\text{dd}}[k,l]\right)$}. Then, the pilot-cancelled received signal is obtained as {\small $\hat{y}_{\text{pc},i,\text{dd}}^{(t)}[k,l] = y_{i,\text{dd}}[k,l] - \hat{y}_{\text{s},i,\text{dd}}^{(t)}[k,l]$}.
The {\small $\hat{y}_{\text{pc},i,\text{dd}}^{(t)}[k,l]$, $i=1,\cdots,n_{r}$} are used to detect the data symbols {\small $x_{\text{d},j}^{(t)}[k,l]$,  $j=1,\cdots,n_{t}$} using the vectorized I/O relation and MMSE-LAS detection. 

\textit{\textbf{Complexity:}}
The complexity of the above scheme is dominated by the computation of MMSE equalizer output vector which involves the inversion of {\small $\mathbf{\hat{H}}_{\tiny\mbox{eff,MIMO}}\mathbf{\hat{H}}^{H}_{\tiny\mbox{eff,MIMO}}$}, whose size is {\small $n_{r}MN\times n_rMN$}. Thus, the order of complexity is {\small $\mathcal{O}(n_{\mathrm{itr}}(n_{r}MN)^{3})$}, where $n_{\mathrm{itr}}$ is the number of turbo iterations.

\begin{figure}
\centering
\includegraphics[width = 8.5cm, height = 6.5cm]{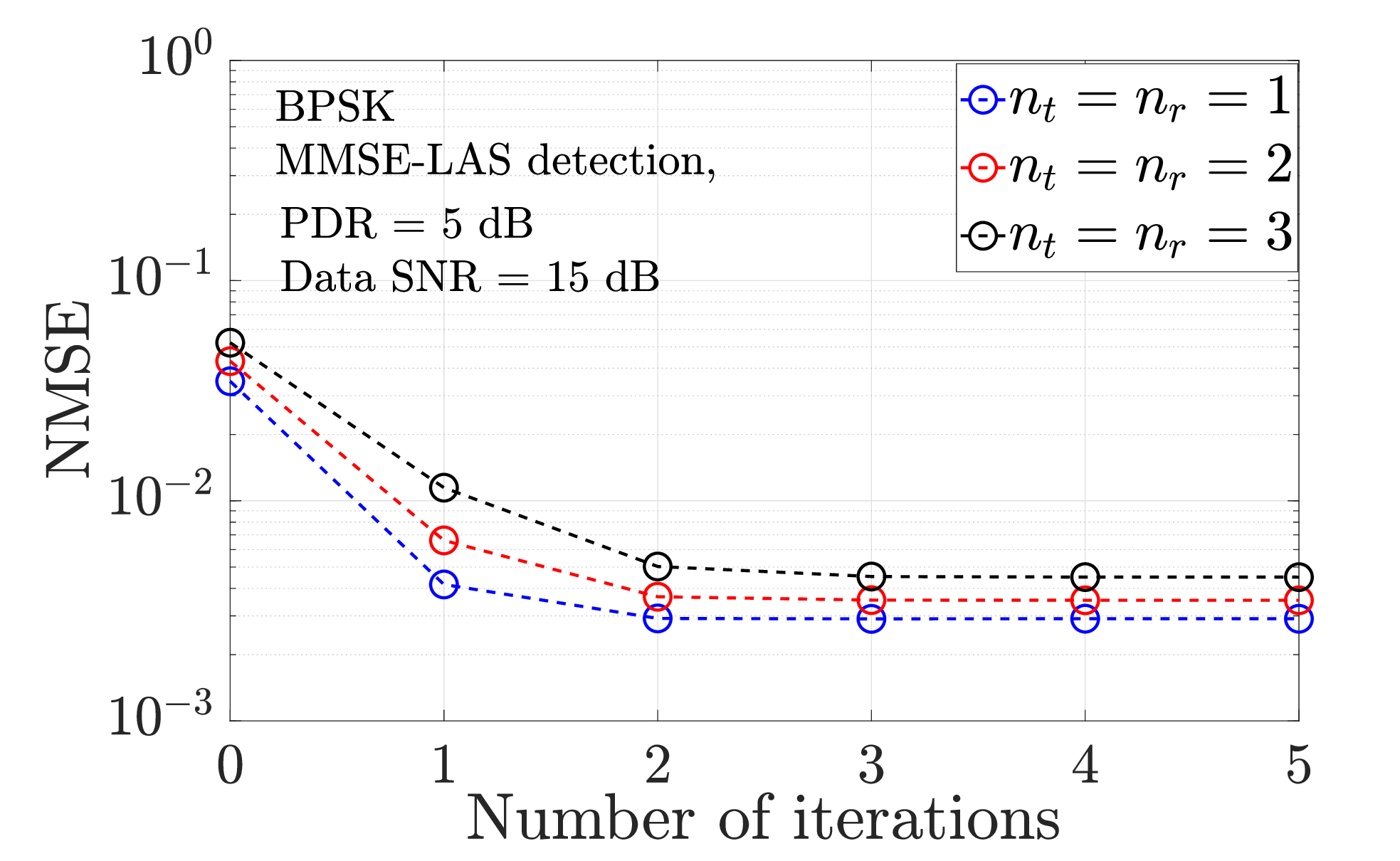}
\caption{NMSE vs number of iterations for $1\times1$, $2\times2$ and $3\times3$ systems at  $5$ dB PDR and $15$ dB data SNR.}
\label{fig:nmse_itr}
\vspace{-2mm}
\end{figure}

\section{Results and discussions}
\label{sec:results}
In this section, we evaluate the performance of the proposed pilot design and estimation/detection scheme in $2\times2$ and $3\times3$ MIMO-Zak-OTFS systems with superimposed spread pilot. Simulation parameters are as follows: Doppler period $\nu_{\text{p}} = 30\,\text{kHz}$, delay period $\tau_{\text{p}} = 1/\nu_{\text{p}} = 33.3\,\mu\text{s}$, $M = 31$, $N=37$, $B= M\nu_{\text{p}}=930 \,\text{kHz}$, $T = 1.233\,\text{ms}$, and BPSK. We consider 
vehicular-A channel model \cite{EVAITU} with fractional DDs, $P=6$ paths, Doppler shift associated with $p\text{th}$ path is $\nu_{\text{max}}\cos{(\theta_{p})}$, where $\theta_{p}\text{s}$ are independent and uniformly distributed in $[0,2\pi)$, $\nu_{\text{max}}=815\,\text{Hz}$, maximum Doppler spread  $2\nu_\text{max} = 1.63\,\text{kHz}$ and maximum delay spread $\tau_\text{max} = 2.51\,\mu$s.
Gaussian-sinc filter at the transmitter and a matching filter at the receive filter, i.e., {\small $w_\text{rx}\left(\tau,\nu\right) = w_\text{tx}^{*}\left(-\tau,-\nu\right)\ e^{j2\pi\tau\nu}$}, are used.

\begin{figure}
\centering
\includegraphics[width = 8.5cm, height = 6.5cm]{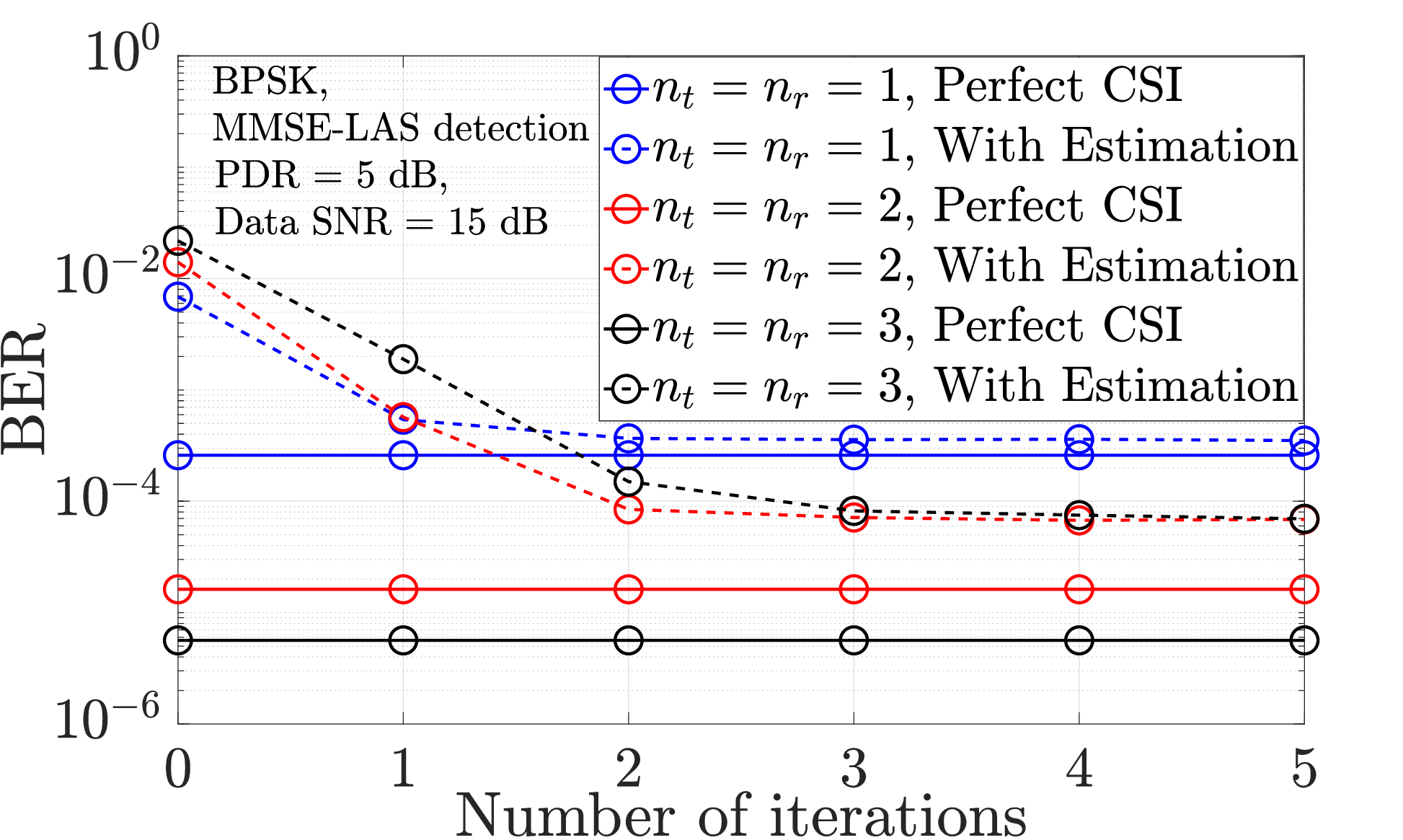}
\vspace{2mm}
\caption{BER vs number of iterations for $1\times1$, $2\times2$ and $3\times3$ systems at  $5$ dB PDR and $15$ dB data SNR.}
\label{fig:ber_itr}
\vspace{-2mm}
\end{figure}

\begin{figure*}[t]
{\small
\begin{eqnarray}
\label{eq:proof_crossambiguity2}
A_{x_{\text{s},2},x_{\text{s},1}}[k,l] & \hspace{-2mm } = & \hspace{-2mm} \frac{1}{MN}\sum\limits_{n_{2}=0}^{N-1}\sum\limits_{n_{1}=0}^{N-1}\sum\limits_{m_{2}=0}^{M-1}\sum\limits_{m_{1}=0}^{M-1}e^{j2\pi\frac{lk}{MN}}e^{j2\pi\frac{(n_{2}l_{\text{p}_{2}}-n_{1}l_{\text{p}_{1}})}{N}} %\\
e^{-j2\pi\frac{(k_{\text{p}_2}m_{2}-k_{\text{p}_{1}}m_{1})}{M}}e^{j2\pi\frac{q_{2}((k_{\text{p}_{2}}+n_{2}M)^{2}+(l_{\text{p}_2}+m_{2}N)^{2})}{MN}} \nonumber \\
& \hspace{-2mm} & \hspace{-2mm} e^{-j2\pi\frac{q_{1}((k+k_{\text{p}_{1}}+n_{1}M)^{2}+(l+l_{\text{p}_1}+m_{1}N)^{2})}{MN}} 
e^{j2\pi\frac{(l+l_{\text{p}_{1}})(k_{\text{p}_{1}+n_{1}M})-l_{\text{p}_{2}}(k_{\text{p}_{2}}+n_{2}M)}{MN}} \nonumber \\
& \hspace{-2mm} & \hspace{-2mm}
\underbrace{\left(\frac{1}{MN}\sum\limits_{k'=0}^{MN-1}e^{-j2\pi\frac{lk'}{MN}}e^{-j2\pi\frac{2k'(q_{2}(k_{\text{p}_{2}}+n_{2}M)-q_{1}(k+k_{\text{p}_{1}}+n_{1}M))}{MN}}
e^{j2\pi\frac{(q_{2}-q_{1})k'^{2}}{MN}}\right)}_{T_1} \nonumber \\
& \hspace{-2mm} & \hspace{-2mm} \underbrace{\left(\frac{1}{MN}\sum\limits_{l'=0}^{MN-1}e^{-j2\pi\frac{2l'(q_{2}(l_{\text{p}_{2}}+m_{2}N)-q_{1}(l+l_{\text{p}_{1}}+m_{1}N))}{MN}}
e^{j2\pi\frac{l'(k_{\text{p}_{2}}+n_{2}M-k_{\text{p}_{1}}-n_{1}M)}{MN}}e^{j2\pi\frac{(q_{2}-q_{1})l'^{2}}{MN}}\right)}_{T_2}.
\end{eqnarray}
}
\hrule
\end{figure*}

In Fig. \ref{fig:nmse_itr}, we present the channel estimation performance in $1\times 1$, $2\times 2$, and $3\times 3$ systems, in terms of normalized mean square error (NMSE), as a function of number of turbo iterations at 5 dB PDR and 15 dB data SNR. NMSE is defined as $\frac{\Sigma_{i=1}^{n_{r}}\Sigma_{j=1}^{n_{t}}\Sigma_{(k,l)\in\mathcal{S}_{o}}|\hat{h}_{\text{eff},i,j}[k,l]-h_{\text{eff},i,j}[k,l]|^{2}}{\Sigma_{i=1}^{n_{r}}\Sigma_{j=1}^{n_{t}}\Sigma_{(k,l)\in\mathcal{S}_{o}}|h_{\text{eff},i,j}[k,l]|^{2}}$. Here, $\mathcal{S}_{o}$ is the support region of the true effective channel. We take \small$\mathcal{S}_{o} = \left\{(k,l)\in\mathbb{Z}\mid -2M+1\leq k \leq 2M-1,\,  -2N+1\leq l \leq 2N-1\right\}$.\\ \normalsize
Also, $\mathcal{S}$, the support region considered for estimation, is a rhombus centered at origin whose diagonal length along delay and Doppler axis is taken to be $16$ and $20$. 
The spread pilots at the transmitters are constructed using point pilot locations as {\small $(k_{\text{p},1},l_{\text{p},1}) = (0,0),\,(k_{\text{p},2},l_{\text{p},2}) = (1,0),\,(k_{\text{p},3},l_{\text{p},3}) = (0,1)$} and slope parameters as $q_{1}=q_{2}=q_{3} =1$. From Fig. \ref{fig:nmse_itr}, it is observed that NMSE decreases as the number of turbo iterations increases, as expected. Also, 3 turbo iterations are found to be adequate to achieve most improvement, beyond which the NMSE floors.

Corresponding to the NMSE performance in Fig. \ref{fig:nmse_itr}, Fig. \ref{fig:ber_itr} shows the corresponding bit error rate (BER) performance as a function of the number of iterations using MMSE-LAS detection. Benefiting from the improvement in NMSE due to turbo iterations, the BER performance also improves as the number of iterations is increased, getting close to the performance with perfect channel state information (CSI). The closeness to perfect CSI performance is compromised for increasing number of antennas due to increased spatial interference. This can be alleviated through the use of better data detection algorithms. 

\begin{figure}
\centering
\includegraphics[width=8.5cm, height=6.5cm]{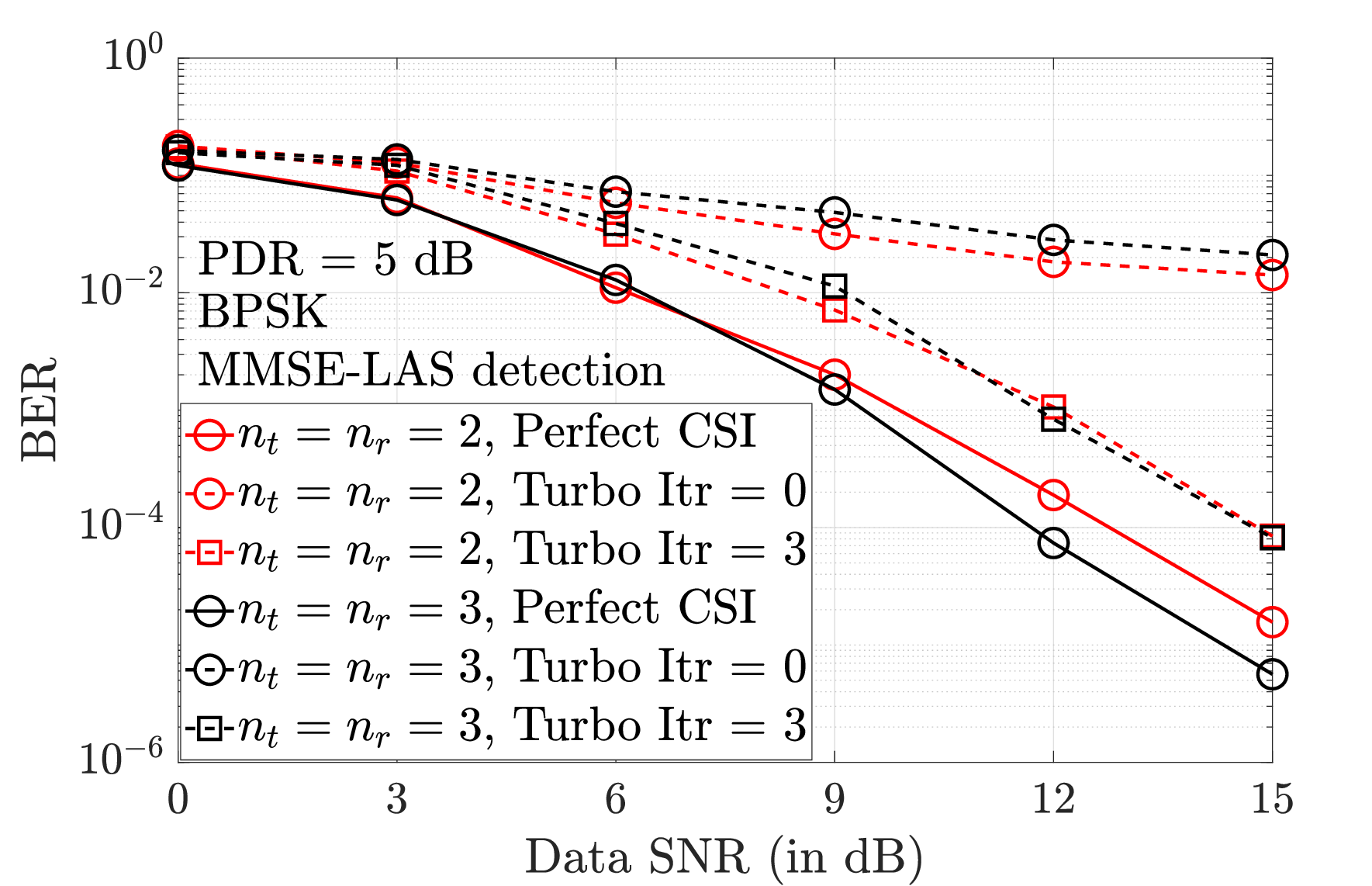}
\caption{BER vs data SNR for $2\times2$ and $3\times3$ MIMO-Zak-OTFS systems at $5$ dB PDR.} 
\label{fig:ber_vs_snr}
\vspace{-2mm}
\end{figure}

Figure \ref{fig:ber_vs_snr} shows the BER performance of $2\times 2$ and $3\times 3$ systems as a function of data SNR at 5 dB PDR. Results for no turbo iterations and 3 iterations are presented. It is observed that significant improvement in BER performance is achieved with 3 turbo iterations, getting close to the performance with perfect CSI, and this illustrates the effectiveness of the proposed pilot design and channel estimation/detection scheme. 

\section{Conclusions}
We considered the problem of pilot sequence design and channel estimation/detection in MIMO-Zak-OTFS with superimposed spread pilots, which has not been reported before. We proposed a pilot design that separates the spread pilots from different transmit antennas in the cross-ambiguity domain at the receiver. This separation allowed channel estimation through a simple read-off operation, which along with turbo iterations between estimation and detection achieved close to perfect CSI performance. Future work could include pilot sequence design for large-scale multiantenna/multiuser systems and learning based transceiver designs.

\appendix
Consider two spread pilot signals ${x_{\text{s},{1},\text{dd}}}[k,l]$ and ${x_{\text{s},{2},\text{dd}}}[k,l]$, where ${x_{\text{s},{1},\text{dd}}}[k,l]$ is obtained by passing a point pilot at $(k_{\text{p}_{1}}$, $l_{\text{p}_{1}})$ through a chirp filter with slope parameter $q_{1}$, and likewise ${x_{\text{s},{2},\text{dd}}}[k,l]$ is obtained with a point pilot at $(k_{\text{p}_{2}}$, $l_{\text{p}_{2}})$ and chirp filter with slope parameter $q_{2}$. The cross-ambiguity between 
{\small ${x_{\text{s},{2},\text{dd}}}[k,l]$} and {\small ${x_{\text{s},{1},\text{dd}}}[k,l]$} is
\begin{eqnarray}
\label{eq:proof_crossambiguity1}
A_{x_{\text{s},2},x_{\text{s},{1}}}[k,l] & \hspace{-2mm } = & \hspace{-2mm} \sum\limits_{k'=0}^{M-1}\sum\limits_{l'=0}^{N-1}x_{\text{s},2,\text{dd}}[k',l']x_{\text{s},{1},\text{dd}}^{*}[k'-k,l'-l] \nonumber \\
&\hspace{-2mm} & \hspace{-2mm} e^{-j2\pi\frac{l(k'-k)}{MN}}.
\end{eqnarray}
Substituting for $x_{\text{s},{2},\text{dd}} [k,l]$ and $x_{\text{s},{1},\text{dd}}[k,l]$ from (\ref{eq:spreadpilot}) in (\ref{eq:proof_crossambiguity1}), using expressions for the chirp filters $w_{2}[k,l]$ and $w_{1}[k,l]$, and applying Lemma 2 in \cite{superimposed_zak}, we obtain
(\ref{eq:proof_crossambiguity2}) given at the top of this page. In the terms $T_{1}$ and $T_{2}$ of (\ref{eq:proof_crossambiguity2}), if $q_{2}-q_{1}\equiv0\pmod{MN} $, then using the fact that sum of all the $(MN){\text{th}}$ roots of unity is zero unless the roots are all unity, we obtain the following two conditions for $A_{x_{\text{s},2},x_{\text{s},{1}}}[k,l]$ to be non-zero:

\vspace{-1mm}
{\footnotesize
\begin{eqnarray}
\label{eq:proof_condition1}
2q_{1}k-l+ M(2n_{1}q_{1}-2n_{2}q_{2})+2q_{1}k_{\text{p}_{1}} -2q_{2}k_{\text{p}_2} & \hspace{-2mm}
\equiv & \nonumber \\
0 \hspace{-1mm} \pmod{MN}, && \\
2q_{1}l_{\text{p}_{1}} - 2q_{2}l_{\text{p}_{2}} + 
N(2q_{1}m_{1} - 
2q_{2}m_{2})+2q_{1}l 
+ k_{\text{p}_{2}}-k_{\text{p}_{1}}
&& \nonumber \\
+ (n_{2}-n_{1})M \equiv 0 \hspace{-1mm} \pmod{MN}. &&
\end{eqnarray}}

\hspace{-4.5mm}
As $q_{1} = q_{2}$ in mod $MN$ arithmetic, we can write the conditions as follows:
\small
\begin{equation}
\label{eq:final_condition1}
2q_{1}(k+k_{\text{p}_{1}}-k_{\text{p}_2})-l\equiv 0 \hspace{-2mm}\pmod{M},
\end{equation}
\begin{equation}
\label{eq:final_conditon2}
\theta_{1}l - 2q_{1}(l_{\text{p}_{1}}-l_{\text{p}_{2}})-k\equiv 0 \hspace{-2mm}\pmod{N},
\end{equation}
\normalsize
where $\theta_{1} = (2q_{1})^{-1} - 2q_{1}$ in mod $MN$ arithmetic. Using (\ref{eq:final_condition1}), (\ref{eq:final_conditon2}) and solving for $k$ and $l$, we obtain
\small
\begin{eqnarray}
\hspace{-8mm}
(1-2q_{1}\theta_{1})
\begin{bmatrix}
\ k    \\
\ l  \\  
\end{bmatrix}
& \hspace{-2.5mm} = & \hspace{-2.5mm}
\underbrace{
\begin{bmatrix}
\ \theta_{1} & 1    \\
\ 1 & 2q_{1} \\  
\end{bmatrix}}_{U}
\begin{bmatrix}
\ nM    \\
\ mN \\ 
\end{bmatrix}
\nonumber \\
& \hspace{-15mm} & \hspace{-10mm}
+\begin{bmatrix}
\ 2q_{1}\theta_{1}(k_{\text{p,1}}-k_{\text{p},2})-2q_{1}(l_{\text{p},1}-l_{\text{p},2})    \\
\ -4q_{1}^{2}(l_{\text{p},1}-l_{\text{p},2})+2q_{1}(k_{\text{p},1}-k_{\text{p},2})\\ 
\end{bmatrix},
\end{eqnarray} 
\normalsize
where $n,m \in \mathbb{Z}.$
This proves that points $(k,l)$ where ambiguity function is non-zero form a lattice obtained by rotating and shifting the period lattice. Also, since $M$, $N$ are odd primes and $q_{1}$ is relatively prime to both, $(1-2q_{1}\theta_{1})$ is not zero in modulo $MN$ arithmetic. Thus, the matrix $U$ is non-singular, and hence there is one-to-one mapping between the shifted lattice and the period lattice. From (\ref{eq:proof_condition1}), we can obtain that for each $n_{2}\in \left\{0,\cdots,N-1\right\}$, there exists a unique $n_{1}\in \left\{0,\cdots,N-1\right\}$ for which the condition is satisfied, given by {\small $n_{1} = n_{2}+\frac{(2q_{1})^{-1}(l-2q_{1}k)-(k_{\text{p}_{1}}-k_{\text{p}_{2}})}{M}\pmod{N}$}. Similarly, {\small $m_{1} = m_{2}+(2q_{1})^{-1}\frac{\theta_{1}l-k-(l_{\text{p}_{1}}-l_{\text{p}_{2}})}{N}\pmod{M}$}. Using the conditions and these unique $n_{1}$ and $m_{1}$ values in (\ref{eq:proof_crossambiguity2}), we can obtain: $|A_{x_{\text{s},{2}},x_{\text{s},{1}}}[k,l]| = 1$. This completes the proof.

\end{document}